\documentclass[ aip,  jmp, singlecolmn]{revtex4-1}
\usepackage{amsmath}
\usepackage{graphicx}
\usepackage{dcolumn}
\usepackage{bm}

\begin{document}

\preprint{}
\title[]{The relationship between the density functional
Hartree-plus-exchange-correlation potential for an integer $N$-electron and $%
\left( N-1\right) $ electron system.}
\author{Daniel P. Joubert}
\email{daniel.joubert2@wits.ac.za}
\affiliation{Centre for Theoretical Physics, University of the Witwatersrand, PO Wits
2050, Johannesburg, South Africa}
\date{\today }

\begin{abstract}
It is shown that the Hartree-plus-exchange-correlation density functional
potential for an integer $N$-electron system differs by a constant form the
corresponding potential for an $\left( N-1\right) $-electron system if the
densities are determined from the same external potential.
\end{abstract}

\maketitle









\section{ Introduction}

In the early 1980's it was discovered that the density functional exchange
correlation potential has a derivative discontinuity when the particle
number crosses and integer\cite{PPLB:82,L.J.Sham1883}. In this paper a
relationship between the potentials of systems with different particle
numbers is proven. It will be shown that the
Hartree-plus-exchange-correlation potential for an integer $N$-electron ($%
N>2)$ system differs by a constant from the corresponding potential of an $%
\left( N-1\right) $-electron system if the densities are determined from the
same external potential. As a corollary it follows that the functional
derivative of the independent particle kinetic energy functional of the $N$
and $\left( N-1\right) $-electron systems also differ by a constant.

\section{Proof}

In the adiabatic connection approach \cite%
{HarrisJones:74,LangrethPerdew:75,LangrethPerdew:77,GunnarsonLundqvist:76}
of the constrained minimization formulation of density functional theory
\cite{HohenbergKohn:64,KohnSham:65,Levy:79,LevyPerdew:85} the Hamiltonian $%
\hat{H}^{\gamma }$ for a system of $N$ electrons is given by
\begin{equation}
\hat{H}_{N}^{\gamma }=\hat{T}^{N}+\gamma \hat{V}_{\text{ee}}^{N}+\hat{v}_{N,%
\text{ext}}^{\gamma }\left[ \rho _{N}\right] .  \label{a3}
\end{equation}%
Atomic units, $\hbar =e=m=1$ are used throughout. $\hat{T}$ is the kinetic
energy operator,%
\begin{equation}
\hat{T}^{N}=-\frac{1}{2}\sum_{i=1}^{N}\nabla _{i}^{2},  \label{a4}
\end{equation}%
\ and $\gamma \hat{V}_{\text{ee }}$is a scaled electron-electron interaction,%
\begin{equation}
\gamma \hat{V}_{\text{ee}}^{N}=\gamma \sum_{i<j}^{N}\frac{1}{\left\vert
\mathbf{r}_{i}-\mathbf{r}_{j}\right\vert }.  \label{a2}
\end{equation}%
The the external potential
\begin{equation}
\hat{v}_{N,\text{ext}}^{\gamma }\left[ \rho _{N}\right] =\sum_{i=1}^{N}v_{%
\text{ext}}^{\gamma }\left( \left[ \rho _{N}\right] ;\mathbf{r}_{i}\right) ,
\label{a1}
\end{equation}%
is constructed to keep the charge density fixed at $\rho _{N}\left( \mathbf{r%
}\right) ,$the ground state charge density of the fully interacting system ($%
\gamma =1$), for all values of the coupling constant $\gamma .$ It has the
form \cite{LevyPerdew:85,GorlingLevy:93}
\begin{align}
v_{\text{ext}}^{\gamma }(\left[ \rho _{N}\right] ;\mathbf{r})& =\left(
1-\gamma \right) v_{ux}([\rho _{N}];\mathbf{r})  \notag \\
& +v_{c}^{1}([\rho _{N}];\mathbf{r)}-v_{c}^{\gamma }([\rho _{N}];\mathbf{r)+}%
v_{\text{ext}}^{1}(\left[ \rho _{N}\right] ;\mathbf{r}),  \label{e1}
\end{align}%
where $v_{\text{ext}}^{1}(\left[ \rho _{N}\right] ;\mathbf{r})=v_{\text{ext}%
}\left( \mathbf{r}\right) $ is the external potential at full coupling
strength, $\gamma =1,$ and $v_{\text{ext}}^{0}(\left[ \rho _{N}\right] ;%
\mathbf{r})$ is non-interacting Kohn-Sham potential. The exchange plus
Hartree potential \cite{ParrYang:bk89,DreizlerGross:bk90}$v_{ux}([\rho _{N}];%
\mathbf{r}),$ is independent of $\gamma ,$ while the correlation potential $%
v_{c}^{\gamma }([\rho _{N}];\mathbf{r)}$ depends in the scaling parameter $%
\gamma .$

\bigskip The chemical potential%
\begin{equation}
\mu =E_{N}^{\gamma }-E_{N-1}^{\gamma }  \label{e2}
\end{equation}%
depends on the asymptotic decay of the charge density \cite%
{ParrYang:bk89,DreizlerGross:bk90,JonesGunnarsson:89}, and hence is
independent of the coupling constant $\gamma $ \cite%
{LevyGorling:96,LevyGorlingb:96}. In Eq. (\ref{e2}) $E_{N-1}^{\gamma }$ is
the groundstate energy of the $\left( N-1\right) $-electron system with the
same single-particle external potential $v_{\text{ext}}^{\gamma }\left( %
\left[ \rho _{N}\right] ;\mathbf{r}\right) $ as the $N$-electron system:%
\begin{eqnarray}
\hat{H}_{N-1}^{\gamma }\left\vert \Psi _{\rho _{N-1}^{\gamma }}^{\gamma
}\right\rangle &=&E_{N-1}^{\gamma }\left\vert \Psi _{\rho _{N-1}^{\gamma
}}^{\gamma }\right\rangle  \notag \\
\hat{H}_{N-1}^{\gamma } &=&\hat{T}^{N-1}+\gamma \hat{V}_{\text{ee}}^{N-1}+%
\hat{v}_{N-1,\text{ext}}^{\gamma }\left[ \rho _{N}\right]  \notag \\
\hat{v}_{N-1,\text{ext}}^{\gamma }\left[ \rho _{N}\right] &=&%
\sum_{i=1}^{N-1}v_{\text{ext}}^{\gamma }\left( \left[ \rho _{N}\right] ;%
\mathbf{r}_{i}\right)  \label{h1}
\end{eqnarray}%
Note that by construction of $v_{\text{ext}}^{\gamma }\left( \left[ \rho _{N}%
\right] ;\mathbf{r}\right) ,$ Eq. (\ref{e1}) $\rho _{N}$ is independent of $%
\gamma ,$ but the groundstate density of the $\left( N-1\right) $-electron
system $\rho _{N-1}^{\gamma }$, is a function of $\gamma .$

The correlation energy $E_{c}^{\gamma }\left[ \rho _{N-1}^{\gamma }\right] $
is defined as\cite{LevyPerdew:85}
\begin{eqnarray}
E_{c}^{\gamma }\left[ \rho _{N-1}^{\gamma }\right] &=&\left\langle \Psi
_{\rho _{N-1}^{\gamma }}^{\gamma }\left\vert \hat{T}^{N-1}+\gamma \hat{V}_{%
\text{ee}}^{N-1}\right\vert \Psi _{\rho _{N-1}^{\gamma }}^{\gamma
}\right\rangle  \notag \\
&&-\left\langle \Psi _{\rho _{N-1}^{\gamma }}^{0}\left\vert \hat{T}%
^{N-1}+\gamma \hat{V}_{\text{ee}}^{N-1}\right\vert \Psi _{\rho
_{N-1}^{\gamma }}^{0}\right\rangle ,  \label{ec1}
\end{eqnarray}%
where $\left\vert \Psi _{\rho _{N-1}^{\gamma }}^{0}\right\rangle $ is the
Kohn-Sham $\left( N-1\right) $ independent particle groundstate wavefunction
that yields the same density as the interacting $\left( N-1\right) $%
-electron system at coupling strength $\gamma .$ Since the correlation part
of the kinetic energy is given by%
\begin{eqnarray}
&&T_{c}^{\gamma }\left[ \rho _{N-1}^{\gamma }\right]  \notag \\
&=&\left\langle \Psi _{\rho _{N-1}^{\gamma }}^{\gamma }\left\vert \hat{T}%
^{N-1}\right\vert \Psi _{\rho _{N-1}^{\gamma }}^{\gamma }\right\rangle
-\left\langle \Psi _{\rho _{N-1}^{\gamma }}^{0}\left\vert \hat{T}%
^{N-1}\right\vert \Psi _{\rho _{N-1}^{\gamma }}^{0}\right\rangle ,
\label{tc}
\end{eqnarray}%
and hence
\begin{eqnarray}
&&\frac{E_{c}^{\gamma }\left[ \rho _{N-1}^{\gamma }\right] -T_{c}^{\gamma }%
\left[ \rho _{N-1}^{\gamma }\right] }{\gamma }  \notag \\
&=&\left\langle \Psi _{\rho _{N-1}^{\gamma }}^{\gamma }\left\vert \hat{V}_{%
\text{ee}}^{N-1}\right\vert \Psi _{\rho _{N-1}^{\gamma }}^{\gamma
}\right\rangle -\left\langle \Psi _{\rho _{N-1}^{\gamma }}^{0}\left\vert
\hat{V}_{\text{ee}}^{N-1}\right\vert \Psi _{\rho _{N-1}^{\gamma
}}^{0}\right\rangle ,  \label{tc1}
\end{eqnarray}%
the derivative of $E_{c}^{\gamma }\left[ \rho _{N-1}^{\gamma }\right] $ with
respect to $\gamma $ can be expressed as \bigskip
\begin{eqnarray}
&&\frac{\partial }{\partial \gamma }E_{c}^{\gamma }\left[ \rho
_{N-1}^{\gamma }\right]  \notag \\
&=&\frac{E_{c}^{\gamma }\left[ \rho _{N-1}^{\gamma }\right] -T_{c}^{\gamma }%
\left[ \rho _{N-1}^{\gamma }\right] }{\gamma }  \notag \\
&&+\left\langle \frac{\partial }{\partial \gamma }\Psi _{\rho _{N-1}^{\gamma
}}^{\gamma }\left\vert \hat{T}^{N-1}+\gamma \hat{V}_{\text{ee}%
}^{N-1}\right\vert \Psi _{\rho _{N-1}^{\gamma }}^{\gamma }\right\rangle
\notag \\
&&+\left\langle \Psi _{\rho _{N-1}^{\gamma }}^{\gamma }\left\vert \hat{T}%
^{N-1}+\gamma \hat{V}_{\text{ee}}^{N-1}\right\vert \frac{\partial }{\partial
\gamma }\Psi _{\rho _{N-1}^{\gamma }}^{\gamma }\right\rangle  \notag \\
&&-\left\langle \frac{\partial }{\partial \gamma }\Psi _{\rho _{N-1}^{\gamma
}}^{0}\left\vert \hat{T}^{N-1}+\gamma \hat{V}_{\text{ee}}^{N-1}\right\vert
\Psi _{\rho _{N-1}^{\gamma }}^{0}\right\rangle  \notag \\
&&-\left\langle \Psi _{\rho _{N-1}^{\gamma }}^{0}\left\vert \hat{T}%
^{N-1}+\gamma \hat{V}_{\text{ee}}^{N-1}\right\vert \frac{\partial }{\partial
\gamma }\Psi _{\rho _{N-1}^{\gamma }}^{0}\right\rangle .  \label{ec2}
\end{eqnarray}%
Upon adding and subtracting (c.c. stands for the complex conjugate of the
previous term)
\begin{eqnarray}
&&\left( \left\langle \frac{\partial }{\partial \gamma }\Psi _{\rho
_{N-1}^{\gamma }}^{\gamma }\left\vert \hat{v}_{N-1,\text{ext}}^{\gamma }%
\left[ \rho _{N}\right] \right\vert \Psi _{\rho _{N-1}^{\gamma }}^{\gamma
}\right\rangle +\text{c.c}\right) +  \notag \\
&&\left( \left\langle \left. \frac{\partial }{\partial \gamma }\Psi _{\rho
_{N-1}^{\gamma }}^{\gamma }\right\vert _{\gamma =0}\left\vert \hat{v}_{N-1,%
\text{ext}}^{0}\left[ \rho _{N}\right] \right\vert \Psi _{\rho
_{N-1}^{0}}^{0}\right\rangle +\text{c.c}\right)  \label{ec3}
\end{eqnarray}%
and utilizing the normalization of the wavefunctions which implies that%
\begin{equation}
\frac{\partial }{\partial \gamma }\left\langle \Psi _{\rho _{N-1}^{\gamma
}}^{\gamma }|\Psi _{\rho _{N-1}^{\gamma }}^{\gamma }\right\rangle =0,
\label{ec4}
\end{equation}%
Eq. (\ref{ec2}) becomes%
\begin{eqnarray}
&&\frac{\partial }{\partial \gamma }E_{c}^{\gamma }\left[ \rho
_{N-1}^{\gamma }\right]  \notag \\
&=&\frac{E_{c}^{\gamma }\left[ \rho _{N-1}^{\gamma }\right] -T_{c}^{\gamma }%
\left[ \rho _{N-1}^{\gamma }\right] }{\gamma }  \notag \\
&&-\left\langle \frac{\partial }{\partial \gamma }\Psi _{\rho _{N-1}^{\gamma
}}^{\gamma }\left\vert \hat{v}_{N-1,\text{ext}}^{\gamma }\left[ \rho _{N}%
\right] \right\vert \Psi _{\rho _{N-1}^{\gamma }}^{\gamma }\right\rangle
-\left\langle \Psi _{\rho _{N-1}^{\gamma }}^{\gamma }\left\vert \hat{v}_{N-1,%
\text{ext}}^{\gamma }\left[ \rho _{N}\right] \right\vert \frac{\partial }{%
\partial \gamma }\Psi _{\rho _{N-1}^{\gamma }}^{\gamma }\right\rangle  \notag
\\
&&+\left\langle \frac{\partial }{\partial \gamma }\Psi _{\rho _{N-1}^{\gamma
}}^{0}\left\vert \hat{v}_{N-1,\text{ext}}^{0}\left[ \rho _{N}\right]
\right\vert \Psi _{\rho _{N-1}^{\gamma }}^{0}\right\rangle +\left\langle
\Psi _{\rho _{N-1}^{\gamma }}^{0}\left\vert \hat{v}_{N-1,\text{ext}}^{0}%
\left[ \rho _{N}\right] \right\vert \frac{\partial }{\partial \gamma }\Psi
_{\rho _{N-1}^{\gamma }}^{0}\right\rangle  \notag \\
&&-\gamma \frac{\partial }{\partial \gamma }\left\langle \Psi _{\rho
_{N-1}^{\gamma }}^{0}\left\vert \hat{V}_{\text{ee}}^{N-1}\right\vert \Psi
_{\rho _{N-1}^{\gamma }}^{0}\right\rangle .  \label{ec5}
\end{eqnarray}%
Now \cite{ParrYang:bk89,DreizlerGross:bk90}
\begin{eqnarray}
\left\langle \Psi _{\rho _{N-1}^{\gamma }}^{0}\left\vert \hat{V}_{\text{ee}%
}^{N-1}\right\vert \Psi _{\rho _{N-1}^{\gamma }}^{0}\right\rangle &=&E_{x}%
\left[ \rho _{N-1}^{\gamma }\right] +U\left[ \rho _{N-1}^{\gamma }\right]
\notag \\
&=&E_{ux}\left[ \rho _{N-1}^{\gamma }\right] .  \label{ux1}
\end{eqnarray}%
is the sum of the exchange $E_{x}\left[ \rho _{N-1}^{\gamma }\right] $ and
mutual Coulomb interaction energy $U\left[ \rho _{N-1}^{\gamma }\right] $ of
the $\left( N-1\right) $-electron system. The charge density $\rho
_{N-1}^{\gamma }$ is a function of $\gamma ,$ therefore \cite{ParrYang:bk89}%
\begin{equation}
\frac{\partial }{\partial \gamma }\left\langle \Psi _{\rho _{N-1}^{\gamma
}}^{0}\left\vert \hat{V}_{\text{ee}}^{N-1}\right\vert \Psi _{\rho
_{N-1}^{\gamma }}^{0}\right\rangle =\int d^{3}r\frac{\partial \rho
_{N-1}^{\gamma }\left( \mathbf{r}\right) }{\partial \gamma }v_{ux}\left( %
\left[ \rho _{N-1}^{\gamma }\right] ;\mathbf{r}\right) ,  \label{ux2}
\end{equation}%
where
\begin{equation}
v_{ux}\left( \left[ \rho _{N-1}^{\gamma }\right] ;\mathbf{r}\right) =\frac{%
\delta }{\delta \rho _{N-1}^{\gamma }\left( \mathbf{r}\right) }\left( E_{x}%
\left[ \rho _{N-1}^{\gamma }\right] +U\left[ \rho _{N-1}^{\gamma }\right]
\right)  \label{ux3}
\end{equation}%
is the sum of the exchange and Hartree potentials for the $\left( N-1\right)
$-electron system. Using Eq. (\ref{ux2}), Eq. (\ref{e1}) and the fact that $%
\left\vert \Psi _{\rho _{N-1}^{\gamma }}^{\gamma }\right\rangle $ and $%
\left\vert \Psi _{\rho _{N-1}^{\gamma }}^{0}\right\rangle $ yield the same
density $\rho _{N-1}^{\gamma }$, Eq. (\ref{ec5}) can be cast as%
\begin{eqnarray}
&&\frac{\partial }{\partial \gamma }E_{c}^{\gamma }\left[ \rho
_{N-1}^{\gamma }\right]  \notag \\
&=&\frac{E_{c}^{\gamma }\left[ \rho _{N-1}^{\gamma }\right] -T_{c}^{\gamma }%
\left[ \rho _{N-1}^{\gamma }\right] }{\gamma }  \notag \\
&&+\int d^{3}r\frac{\partial \rho _{N-1}^{\gamma }\left( \mathbf{r}\right) }{%
\partial \gamma }\left( v_{c}^{\gamma }\left( \left[ \rho _{N}\right] ;%
\mathbf{r}\right) +\gamma v_{ux}\left( \left[ \rho _{N}\right] ;\mathbf{r}%
\right) -\gamma v_{ux}\left( \left[ \rho _{N-1}^{\gamma }\right] ;\mathbf{r}%
\right) \right)  \label{ec6}
\end{eqnarray}

From the definition of $E_{c}^{\gamma }\left[ \rho _{N-1}^{\gamma }\right] ,$
Eq. (\ref{ec1}), the correlation energy can also be written as
\begin{equation}
E_{c}^{\gamma }\left[ \rho _{N-1}^{\gamma }\right] =\hat{T}^{\gamma }\left[
\rho _{N-1}^{\gamma }\right] -\hat{T}^{0}\left[ \rho _{N-1}^{\gamma }\right]
+\gamma \hat{V}_{\text{ee}}\left[ \rho _{N-1}^{\gamma }\right] -\gamma E_{ux}%
\left[ \rho _{N-1}^{\gamma }\right]  \label{ec7}
\end{equation}%
where
\begin{eqnarray}
\hat{T}^{\gamma }\left[ \rho _{N-1}^{\gamma }\right] &=&\left\langle \Psi
_{\rho _{N-1}^{\gamma }}^{\gamma }\left\vert \hat{T}^{N-1}\right\vert \Psi
_{\rho _{N-1}^{\gamma }}^{\gamma }\right\rangle  \notag \\
\hat{T}^{0}\left[ \rho _{N-1}^{\gamma }\right] &=&\left\langle \Psi _{\rho
_{N-1}^{\gamma }}^{0}\left\vert \hat{T}^{N-1}\right\vert \Psi _{\rho
_{N-1}^{\gamma }}^{0}\right\rangle  \label{ec8} \\
.\hat{V}_{\text{ee}}\left[ \rho _{N-1}^{\gamma }\right] &=&\left\langle \Psi
_{\rho _{N-1}^{\gamma }}^{\gamma }\left\vert \hat{V}_{\text{ee}%
}^{N-1}\right\vert \Psi _{\rho _{N-1}^{\gamma }}^{\gamma }\right\rangle .
\end{eqnarray}%
It now follows that%
\begin{eqnarray}
&&\frac{\partial }{\partial \gamma }E_{c}^{\gamma }\left[ \rho
_{N-1}^{\gamma }\right]  \notag \\
&=&\frac{E_{c}^{\gamma }\left[ \rho _{N-1}^{\gamma }\right] -T_{c}^{\gamma }%
\left[ \rho _{N-1}^{\gamma }\right] }{\gamma }  \notag \\
&&+\int d^{3}r\frac{\partial \rho _{N-1}^{\gamma }\left( \mathbf{r}\right) }{%
\partial \gamma }\left( \frac{\delta T_{N-1}^{\gamma }\left[ \rho
_{N-1}^{\gamma }\right] }{\delta \rho _{N-1}^{\gamma }\left( \mathbf{r}%
\right) }-\frac{\delta T_{N-1}^{0}\left[ \rho _{N-1}^{\gamma }\right] }{%
\delta \rho _{N-1}^{\gamma }\left( \mathbf{r}\right) }-\gamma \frac{\delta
E_{ux}\left[ \rho _{N-1}^{\gamma }\right] }{\delta \rho _{N-1}^{\gamma
}\left( \mathbf{r}\right) }\right)  \notag \\
&&+\gamma \frac{\partial }{\partial \gamma }V_{\text{ee}}\left[ \rho
_{N-1}^{\gamma }\right]  \label{ec9}
\end{eqnarray}%
where use was made of the relation\bigskip
\begin{equation}
\frac{E_{c}^{\gamma }\left[ \rho _{N-1}^{\gamma }\right] -T_{c}^{\gamma }%
\left[ \rho _{N-1}^{\gamma }\right] }{\gamma }=\hat{V}_{\text{ee}}\left[
\rho _{N-1}^{\gamma }\right] -E_{ux}\left[ \rho _{N-1}^{\gamma }\right] .
\label{ec10}
\end{equation}%
The last term in Eq. (\ref{ec9}) can be transformed as follows:%
\begin{eqnarray}
&&\frac{\partial }{\partial \gamma }V_{\text{ee}}\left[ \rho _{N-1}^{\gamma }%
\right]  \notag \\
&=&\left\langle \frac{\partial }{\partial \gamma }\Psi _{\rho _{N-1}^{\gamma
}}^{\gamma }\left\vert \hat{V}_{\text{ee}}^{N-1}\right\vert \Psi _{\rho
_{N-1}^{\gamma }}^{\gamma }\right\rangle +\left\langle \Psi _{\rho
_{N-1}^{\gamma }}^{\gamma }\left\vert \hat{V}_{\text{ee}}^{N-1}\right\vert
\frac{\partial }{\partial \gamma }\Psi _{\rho _{N-1}^{\gamma }}^{\gamma
}\right\rangle  \notag \\
&=&\frac{1}{\gamma }\left( \left\langle \frac{\partial }{\partial \gamma }%
\Psi _{\rho _{N-1}^{\gamma }}^{\gamma }\left\vert \hat{T}^{N-1}+\gamma \hat{V%
}_{ee}^{N-1}+\hat{v}_{N-1,\text{ext}}^{\gamma }\left[ \rho _{N}\right]
\right\vert \Psi _{\rho _{N-1}^{\gamma }}^{\gamma }\right\rangle \right.
\notag \\
&&+\left. \left\langle \Psi _{\rho _{N-1}^{\gamma }}^{\gamma }\left\vert
\hat{T}^{N-1}+\gamma \hat{V}_{ee}^{N-1}+\hat{v}_{N-1,\text{ext}}^{\gamma }%
\left[ \rho _{N}\right] \right\vert \frac{\partial }{\partial \gamma }\Psi
_{\rho _{N-1}^{\gamma }}^{\gamma }\right\rangle \right)  \notag \\
&&-\frac{1}{\gamma }\left( \left\langle \frac{\partial }{\partial \gamma }%
\Psi _{\rho _{N-1}^{\gamma }}^{\gamma }\left\vert \hat{T}^{N-1}+\hat{v}_{N-1,%
\text{ext}}^{\gamma }\left[ \rho _{N}\right] \right\vert \Psi _{\rho
_{N-1}^{\gamma }}^{\gamma }\right\rangle \right.  \notag \\
&&+\left. \left\langle \Psi _{\rho _{N-1}^{\gamma }}^{\gamma }\left\vert
\hat{T}^{N-1}+\hat{v}_{N-1,\text{ext}}^{\gamma }\left[ \rho _{N}\right]
\right\vert \frac{\partial }{\partial \gamma }\Psi _{\rho _{N-1}^{\gamma
}}^{\gamma }\right\rangle \right)  \notag \\
&=&-\frac{1}{\gamma }\int d^{3}r\frac{\partial \rho _{N-1}^{\gamma }\left(
\mathbf{r}\right) }{\partial \gamma }\left( \frac{\delta T_{N-1}^{\gamma }%
\left[ \rho _{N-1}^{\gamma }\right] }{\delta \rho _{N-1}^{\gamma }\left(
\mathbf{r}\right) }+v_{\text{ext}}^{\gamma }(\left[ \rho _{N}\right] ;%
\mathbf{r})\right) .  \label{ec11}
\end{eqnarray}%
Here use was made Eqs. (\ref{h1}) and (\ref{ec4}). Combining Eqs. (\ref{ec9}%
) and (\ref{ec11}) leads to \bigskip
\begin{eqnarray}
&&\frac{\partial }{\partial \gamma }E_{c}^{\gamma }\left[ \rho
_{N-1}^{\gamma }\right]  \notag \\
&=&\frac{E_{c}^{\gamma }\left[ \rho _{N-1}^{\gamma }\right] -T_{c}^{\gamma }%
\left[ \rho _{N-1}^{\gamma }\right] }{\gamma }  \notag \\
&&+\int d^{3}r\frac{\partial \rho _{N-1}^{\gamma }\left( \mathbf{r}\right) }{%
\partial \gamma }\left( -v_{\text{ext}}^{\gamma }(\left[ \rho _{N}\right] ;%
\mathbf{r})-\frac{\delta T_{N-1}^{0}\left[ \rho _{N-1}^{\gamma }\right] }{%
\delta \rho _{N-1}^{\gamma }\left( \mathbf{r}\right) }-\gamma v_{ux}\left( %
\left[ \rho _{N-1}^{\gamma }\right] ;\mathbf{r}\right) \right) .
\label{ec12a}
\end{eqnarray}%
At the solution point\cite{ParrYang:bk89,DreizlerGross:bk90} the following
equation has to be satisfied for the $\left( N-1\right) $-electron system:
\begin{equation}
\frac{\delta T_{N-1}^{0}\left[ \rho _{N-1}^{\gamma }\right] }{\delta \rho
_{N-1}^{\gamma }\left( \mathbf{r}\right) }+\gamma v_{ux}\left( \left[ \rho
_{N-1}^{\gamma }\right] ;\mathbf{r}\right) +v_{c}^{\gamma }\left( \left[
\rho _{N-1}^{\gamma }\right] ;\mathbf{r}\right) +v_{\text{ext}}^{\gamma }(%
\left[ \rho _{N}\right] ;\mathbf{r})=\mu _{N-1}^{\gamma },  \label{s1}
\end{equation}%
where $\mu _{N-1}^{\gamma }$ is a constant, the chemical potential of the $%
\left( N-1\right) $-electrons moving in the external potential $v_{\text{ext}%
}^{\gamma }(\left[ \rho _{N}\right] ;\mathbf{r}).$ Taking into account that
the number of electrons is fixed at $\left( N-1\right) $ independent of $%
\gamma ,$ it follows that
\begin{eqnarray}
\int d^{3}r\frac{\partial \rho _{N-1}^{\gamma }\left( \mathbf{r}\right) }{%
\partial \gamma } &=&\frac{\partial }{\partial \gamma }\left( N-1\right)
\notag \\
&=&0,  \label{p1}
\end{eqnarray}%
where it is assumed that the order of the integration and derivative can be
reversed. From Eqs. (\ref{ec12a}), (\ref{s1}) and (\ref{p1})
\begin{eqnarray}
&&\frac{\partial }{\partial \gamma }E_{c}^{\gamma }\left[ \rho
_{N-1}^{\gamma }\right]  \notag \\
&=&\frac{E_{c}^{\gamma }\left[ \rho _{N-1}^{\gamma }\right] -T_{c}^{\gamma }%
\left[ \rho _{N-1}^{\gamma }\right] }{\gamma }+\int d^{3}r\frac{\partial
\rho _{N-1}^{\gamma }\left( \mathbf{r}\right) }{\partial \gamma }%
v_{c}^{\gamma }\left( \left[ \rho _{N-1}^{\gamma }\right] ;\mathbf{r}\right)
.  \label{ec13}
\end{eqnarray}%
Comparing Eqs. (\ref{ec6}) and (\ref{ec13}) shows that%
\begin{equation}
0=\int d^{3}r\frac{\partial \rho _{N-1}^{\gamma }\left( \mathbf{r}\right) }{%
\partial \gamma }\left( v_{uxc}^{\gamma }\left( \left[ \rho _{N}\right] ;%
\mathbf{r}\right) -v_{uxc}^{\gamma }\left( \left[ \rho _{N-1}^{\gamma }%
\right] ;\mathbf{r}\right) \right)  \label{ec14}
\end{equation}%
where
\begin{equation}
v_{uxc}^{\gamma }\left( \left[ \rho \right] ;\mathbf{r}\right)
=v_{c}^{\gamma }\left( \left[ \rho \right] ;\mathbf{r}\right) +\gamma
v_{ux}\left( \left[ \rho \right] ;\mathbf{r}\right)  \label{uxc1}
\end{equation}%
is the Hartree plus exchange-correlation potential for a system with density
$\rho $ and coupling strength $\gamma .$

The charge density $\rho _{N-1}^{\gamma }\left( \mathbf{r}\right) $ is a
functional of the potential $v_{\text{ext}}^{\gamma }\left( \left[ \rho _{N}%
\right] ;\mathbf{r}\right) $ as can be seen from Eq. (\ref{h1}) \cite%
{ParrYang:bk89,DreizlerGross:bk90}. Therefore%
\begin{eqnarray}
\frac{\partial \rho _{N-1}^{\gamma }\left( \mathbf{r}\right) }{\partial
\gamma } &=&\int d^{3}r^{\prime }\left. \frac{\delta \rho _{N-1}^{\gamma
}\left( \mathbf{r}\right) }{\delta v_{\text{ext}}^{\gamma }\left( \left[
\rho _{N}\right] ;\mathbf{r}^{\prime }\right) }\right\vert _{N-1}\frac{%
\partial }{\partial \gamma }v_{\text{ext}}^{\gamma }\left( \left[ \rho _{N}%
\right] ;\mathbf{r}^{\prime }\right)  \notag \\
&=&-\int d^{3}r^{\prime }\left. \frac{\delta \rho _{N-1}^{\gamma }\left(
\mathbf{r}\right) }{\delta v_{\text{ext}}^{\gamma }\left( \left[ \rho _{N}%
\right] ;\mathbf{r}^{\prime }\right) }\right\vert _{N-1}\left( v_{ux}([\rho
_{N}];\mathbf{r}^{\prime })+\frac{\partial }{\partial \gamma }v_{c}^{\gamma
}([\rho _{N}];\mathbf{r}^{\prime }\mathbf{)}\right) .
\end{eqnarray}%
Now%
\begin{equation}
\left. \frac{\delta \rho _{N-1}^{\gamma }\left( \mathbf{r}\right) }{\delta
v_{\text{ext}}^{\gamma }\left( \left[ \rho _{N}\right] ;\mathbf{r}^{\prime
}\right) }\right\vert _{N-1}=\chi _{N-1}^{\gamma }\left( \mathbf{r},\mathbf{r%
}^{\prime }\right)  \label{drs1}
\end{equation}%
is the density response function of the $\left( N-1\right) $-particle
system. From stability considerations $\chi _{N-1}^{\gamma }\left( \mathbf{r}%
,\mathbf{r}^{\prime }\right) $ is negative semi-definite and has one zero
eigenvalue which corresponds to the invariance of the density when the
potential is changed by a constant \cite{ParrYang:bk89,DreizlerGross:bk90}.
This implies that
\begin{eqnarray}
0 &=&\int d^{3}r\frac{\partial \rho _{N-1}^{\gamma }\left( \mathbf{r}\right)
}{\partial \gamma }f\left( \mathbf{r}\right)  \notag \\
&=&-\int d^{3}r\int d^{3}r^{\prime }f\left( \mathbf{r}\right) \chi
_{N-1}^{\gamma }\left( \mathbf{r},\mathbf{r}^{\prime }\right) \left(
v_{ux}([\rho _{N}];\mathbf{r}^{\prime })+\frac{\partial }{\partial \gamma }%
v_{c}^{\gamma }([\rho _{N}];\mathbf{r}^{\prime }\mathbf{)}\right)
\label{drs2}
\end{eqnarray}%
is only possible if $f\left( \mathbf{r}\right) =$ constant since $%
v_{ux}([\rho _{N}];\mathbf{r}^{\prime })+\frac{\partial }{\partial \gamma }%
v_{c}^{\gamma }\left( [\rho _{N}];\mathbf{r}^{\prime }\right) \neq $
constant. This proves the main point of this paper:%
\begin{equation}
v_{uxc}^{\gamma }\left( \left[ \rho _{N}\right] ;\mathbf{r}\right)
=v_{uxc}^{\gamma }\left( \left[ \rho _{N-1}^{\gamma }\right] ;\mathbf{r}%
\right) +\text{constant}  \label{proof}
\end{equation}%
when $\rho _{N}$ and $\rho _{N-1}^{\gamma }$ are determined by the same
external potential $v_{\text{ext}}^{\gamma }\left( \left[ \rho _{N}\right] ;%
\mathbf{r}\right) .$

\section{Proof of corollary}

Let the energy functional $F_{N}^{\gamma }$\bigskip $\left[ \rho _{N}\right]
$ \cite{ParrYang:bk89,DreizlerGross:bk90} be defined as%
\begin{equation}
F_{N}^{\gamma }\left[ \rho _{N}\right] =T_{N}^{0}\left[ \rho _{N}\right]
+\gamma U\left[ \rho _{N}\right] +\gamma E_{x}\left[ \rho _{N}\right]
+E_{c}^{\gamma }\left[ \rho _{N}\right] .  \label{f1}
\end{equation}%
Then \cite{ParrYang:bk89,DreizlerGross:bk90}%
\begin{equation}
\frac{\delta F_{N}^{\gamma }\left[ \rho _{N}\right] }{\delta \rho _{N}\left(
\mathbf{r}\right) }+v_{\text{ext}}^{\gamma }(\left[ \rho _{N}\right] ;%
\mathbf{r})=\mu  \label{f2}
\end{equation}%
and

\bigskip
\begin{equation}
\frac{\delta F_{N}^{\gamma }\left[ \rho _{N-1}^{\gamma }\right] }{\delta
\rho _{N-1}^{\gamma }\left( \mathbf{r}\right) }+v_{\text{ext}}^{\gamma }(%
\left[ \rho _{N}\right] ;\mathbf{r})=\mu _{N-1}^{\gamma }.  \label{f3}
\end{equation}%
From the last two equations we find that%
\begin{equation}
\frac{\delta F_{N}^{\gamma }\left[ \rho _{N}\right] }{\delta \rho _{N}\left(
\mathbf{r}\right) }-\frac{\delta F_{N}^{\gamma }\left[ \rho _{N-1}^{\gamma }%
\right] }{\delta \rho _{N-1}^{\gamma }\left( \mathbf{r}\right) }=\mu -\mu
_{N-1}^{\gamma }  \label{f4}
\end{equation}%
and therefore, using Eqs. (\ref{proof}) and (\ref{f1}), it follows that
\begin{equation}
\frac{\delta T_{N}^{0}\left[ \rho _{N}\right] }{\delta \rho _{N}\left(
\mathbf{r}\right) }-\frac{\delta T_{N}^{0}\left[ \rho _{N-1}^{\gamma }\right]
}{\delta \rho _{N-1}^{\gamma }\left( \mathbf{r}\right) }=\text{constant}
\label{f5}
\end{equation}

\section{Discussion and summary}

The relationship in Eq. (\ref{proof}) is valid for $N>2.$ This follows from
the step in Eq. (\ref{ec13}) where the correlation energy of the $N-1$
electron system is taken as non-zero. In the proof use is made of the $N$
and $N-1$ electron wave functions, hence the proof given here is valid for
integer $N.$ The proof for non-integer values of the electrons and will be
presented in another paper.

In summary, it was shown that the Hartree-plus-exchange-correlation
potential for an integer $N$-electron ($N>2)$ system differs by a constant
form the corresponding potential for an $\left( N-1\right) $-electron system
if the densities are determined with the same external potential. As a
corollary it was shown that the functional derivative of the independent
particle kinetic energy functional of the $N$ and $\left( N-1\right) $%
-electron systems also differ by a constant.

\end{document}